\documentclass[prd,twocolumn,showpacs,preprintnumbers,amsmath,amssymb,superscriptaddress,floatfix,nofootinbib]{revtex4-2}

\usepackage{graphicx}
\usepackage{bm}
\usepackage{amsmath}
\usepackage{amsfonts}
\usepackage{amssymb}
\usepackage{color}
\usepackage{multirow}
\usepackage{subfigure}
\usepackage{txfonts}
\usepackage[colorlinks, citecolor=blue,anchorcolor=red,menucolor=red, linkcolor=red,filecolor=red,runcolor=red,urlcolor=blue,frenchlinks=red]{hyperref}

\begin{document}

%\title{Theoretical study of $N(1535)$ and $a_0(980)$ in the process $\Lambda_c^+ \to \pi^+\eta n$}
\title{Roles of the $N(1535)$ and $a_0(980)$ in the process $\Lambda_c^+ \to \pi^+\eta n$}

\date{\today}
\author{Meng-Yuan Li}
\affiliation{School of Physics, Zhengzhou
	University, Zhengzhou 450001, China}

\author{Wen-Tao Lyu}\email{lvwentao9712@163.com}
\affiliation{School of Physics, Zhengzhou
	University, Zhengzhou 450001, China}
	
\author{Li-Juan Liu}\email{liulijuan@zzu.edu.cn}
\affiliation{School of Physics, Zhengzhou
	University, Zhengzhou 450001, China}
	
\author{En Wang}\email{wangen@zzu.edu.cn}
\affiliation{School of Physics, Zhengzhou
	University, Zhengzhou 450001, China}

\begin{abstract}

%We have investigated the process $\Lambda_c^+ \to \pi^+\eta n$ by taking into account the $S$-wave pseudosalar meson-octet baryon interactions and $S$-wave pseudosalar-pseudosalar interactions within the chiral unitary approach, where the $N(1535)$ and $a_0(980)$ could be dynamically generated. Our results show that the $\eta n$ invariant mass distribution has a significant peak structure associated to $N(1535)$. We have also calculated the $\pi^{+}\eta$ invariant mass distribution and found that there is a cusp structure corresponding to $a_0(980)$. For the purpose of presenting the contribution of the $a_0(980)$ more clearly, we performed a cut and removed the part where $M_{\eta n}$ is below 1700 MeV. We have also roughly estimated the ratio $R$ = $\mathcal{B}(\Lambda_c^+ \to a_0(980)^{+} n)/\mathcal{B}(\Lambda_c^+ \to \pi^+\eta n)\approx 0.313$. We encourage our experimental colleagues to search for the resonance $a_0(980)$ and $N(1535)$ signal in the process $\Lambda_c^+ \to \pi^+\eta n$.

We have investigated the process $\Lambda_c^+ \to \pi^+\eta n$ by taking into account the contributions from the nucleon resonance $N(1535)$ and the scalar meson $a_0(980)$, which could be dynamically generated by the interaction of the $S$-wave pseudosalar meson-octet baryon and the $S$-wave pseudosalar meson-pseudosalar meson, respectively. Our results show that, in $\eta n$ invariant mass distribution, there is a significant near-threshold enhancement structure, which could be associated with $N(1535)$. On the other hand, one can find a clear cusp structure of $a_0(980)$ in $\pi^+\eta$ invariant mass distribution. We further estimate the ratio $R$ = $\mathcal{B}(\Lambda_c^+ \to a_0(980)^+ n)/\mathcal{B}(\Lambda_c^+ \to \pi^+\eta n)\approx 0.313$. Our results can be tested by BESIII,  Belle~II, and the proposed Super Tau-Charm Facility experiments in the future.

\end{abstract}

\maketitle

%%%%%%%%%%%%%%%%%%%%%%
\section{Introduction} \label{sec:Introduction}

Since $X(3872)$ was observed by the Belle Collaboration in 2003~\cite{Belle:2003nnu}, many candidates of the exotic states were reported by experiments, and called many attentions~\cite{Chen:2022asf,Brambilla:2019esw,Meng:2022ozq,Guo:2023xyf,Liu:2024uxn}. In the light baryons region, there are also some puzzles, and one of these puzzles is the mass reverse problem, i.e. the mass of $N(1535)$ with negative parity should be smaller than that of the radial excited state $N(1440)$ with the spin-parity quantum numbers of $J^P=1/2^+$, but the experimental observations are just the opposite~\cite{ParticleDataGroup:2024cfk}. Moreover, within traditional quark model, the $N(1535)$ does not contain the $s\bar{s}$ component, however it has a strong coupling effect with $\eta N$ and $K\Lambda$ channels, which implies that the $N(1535)$ should have sizeable pentaquark component with hidden strangeness of $s\bar{s}$~\cite{Liu:2005pm,Xie:2017erh,Pavao:2018wdf}. 

In Refs.~\cite{Helminen:2000jb,Zou:2007mk,Zhang:2004xt,Hannelius:2000gu}, the authors deem that the $N(1535)$ resonance could be the lowest $L=1$ orbital excited $uud$ state with a large admixing of the $[ud][us]\bar{s}$ pentaquark component, which leads to the $N(1535)$ has a heavier mass than the $N(1440)$, and also gives a natural explanation for the large couplings of $N(1535)$ to the strangeness channels $\eta N$ and $K\Lambda$. In addition, the $N(1535)$ state could be dynamically generated via $S$-wave pseudoscalar meson-octet baryon interaction within the chiral unitary approach, and is predicted to strongly couple to the channels $\pi N$, $\eta N$, $K\Lambda$, and $K\Sigma$~\cite{Kaiser:1995eg,Kaiser:1996js,Inoue:2001ip,Bruns:2010sv,Nieves:2011gb,Gamermann:2011mq,Khemchandani:2013nma}. When the pseudoscalar meson-baryon mixing with the vector meson-baryon states was considered, the physical picture remains unchanged~\cite{Garzon:2014ida}. Furthermore, using above approach, its mass and width obtained from the position of the pole on the second Riemann sheet are in line with the experimental results~\cite{Nieves:2001wt,Nieves:2011gb,Gamermann:2011mq,Khemchandani:2013nma,Garzon:2014ida,Inoue:2001ip,Bruns:2010sv}. In Refs.~\cite{Xie:2007qt,Doring:2008sv}, their analyses of the near-threshold  $\phi$ production in $\pi p$ and $pp$ collisions suggest that the $N(1535)$ strongly couples to the $K\Sigma$, $K\Lambda$ and $\phi N$ channels, which is consistent with the results of the former approach. Then, in Refs.~\cite{Abell:2023nex,Guo:2022hud,Liu:2015ktc}, the $N(1535)$ was also interpreted as a three-quark core dressed by meson-nucleon scattering contributions within the Hamiltonian effective field theory. Recently, it was suggested to test the molecular nature of $N(1535)$ by measuring its correlation functions~\cite{Molina:2023jov}, or the scattering length and effective range of the channels $K\Sigma$, $K\Lambda$, and $\eta p$~\cite{Li:2023pjx}. Thus, the properties of $N(1535)$ with the spin-parity quantum numbers of $J^{P}=1/2^-$ still need to be further explored at present~\cite{Klempt:2007cp,Crede:2013kia}.

On the other hand, as we know, the identification of the light scalar mesons is very difficult, resulting from their large decay widths, and the structures of the scalar mesons have many interpretations, i.e. traditional $q\bar{q}$ states, multiquark states, hadronic molecules, or glueballs, etc~\cite{Close:2002zu,Amsler:2004ps,Bugg:2004xu,Klempt:2007cp,Pelaez:2015qba}. Among many light scalar mesons, the light scalar meson $a_0(980)$ also has been explained to be either a molecular state, a tetraquark state, a conventional $q\bar{q}$ meson, or the mixing of different components~\cite{Klempt:2007cp,Nieves:1998hp,Janssen:1994wn,Wolkanowski:2015lsa}. In the chiral unitary approach, $a_0(980)$ could be dynamically generated from the $S$-wave interaction of the coupled channels $K\bar{K}$ and $\pi\eta$~\cite{Oller:1997ti,Nieves:1998hp}, which has been widely used in many theoretical studies~\cite{Wang:2020pem,Xie:2014tma,Oset:2016lyh,Duan:2020vye,Zhu:2022guw,Feng:2020jvp,Wang:2022nac,Wang:2021naf,Liang:2016hmr}.

%for the light scalar mesons like $a_0(980)$, $f_0(500)$ and $f_0(980)$, which have relatively low masses and the same quantum numbers as vacuum $J^{PC}=0^{++}$, their masses fail to consistent with the predictions of the quark model~\cite{ParticleDataGroup:2022pth,tHooft:2008rus,Deng:2012wj}.

The non-leptonic weak decays of charmed baryons are an important way to explore the properties of these light hadrons, since those processes have large phase space and involve complicated final-state interactions
~\cite{Oset:2016lyh,Miyahara:2015cja, Hyodo:2011js,Wang:2022nac,Wang:2020pem,Zeng:2020och,Feng:2020jvp,Li:2024rqb,Zhang:2024jby,Wang:2024jyk}. 
For instance, based on the suggestion of Refs.~\cite{Xie:2017xwx,Lyu:2024qgc}, the BESIII Collaboration has observed the evidence of the $\Sigma(1/2^-)$ with mass of 1380~MeV and statistical significance $>3\sigma$ in the process $\Lambda_c^+\to \Lambda \pi^+\eta$, and reported a branching
fraction of 54\% for the $\Lambda a_0(980)$ decay mode~\cite{BESIII:2024mbf}. However, Ref.~\cite{Duan:2024led} has well reproduced the BESIII measured invariant mass distributions of this process by considering the contributions from the $a_0(980)$, $\Lambda(1670)$, and $\Sigma(1385)$, and found the branching ratio for the $\Lambda_c^+\to\Lambda a_0(980)$ decay mode is about one half of that obtained by BESIII. 
The reaction of $\Lambda_c^+\to \Lambda \pi^+\eta$ measured by BESIII could be related to the $\Lambda_c^+\to \pi^+\eta n$~\cite{BESIII:2024mbf}, since the former one is Cabibbo-favored and the latter one is Cabibbo-suppressed. It it interesting to make theoretical prediction for the latter process $\Lambda_c^+\to \pi^+\eta n$, and the future measurements of this process with high statistic experiments could shed light on the production mechanism of the $a_0(980)$ in the charmed baryon decays.
Thus, we propose to study the similar process $\Lambda_c^+\to \pi^+\eta n$, which could be used to understand the properties of the $N(1535)$ and $a_0(980)$, and also to solve this puzzle.

In this work, we will analyze the process $\Lambda_c^+ \to \pi^+\eta n$ using the chiral unitary approach considering the $S$-wave pseudosalar meson-octet baryon interaction, which will dynamically generate the resonance $N(1535)$. Moreover, we also taking into account the contribution of scalar meson $a_0(980)$ from the $S$-wave pseudosalar meson-pseudosalar meson interaction. Then, we calculate the $\eta n$ and $\pi^+\eta$ invariant mass distributions in the $\Lambda_c^+ \to \pi^+\eta n$ reaction. The work done here should be an incentive for this more accurate experimental analysis to be performed.

The paper is organized as follows. In Sec.~\ref{sec:Formalism}, we show the
theoretical formalism for the process $\Lambda_c^+ \to \pi^+\eta n$. Our numerical results and discussions are presented in Sec.~\ref{sec:Results}. A brief summary of this work is provided in the last section.
%%%%%%%%%%%%%%%%%%%%%%
\section{Formalism} \label{sec:Formalism}

In this section, we present the theoretical formalism for the process $\Lambda_c^+ \to \pi^+\eta n$. We first demonstrate the $N(1535)$ is dynamically generated by the $S$-wave pseudoscalar meson-octet baryon interaction in Sec.~\ref{sec2a}. Then, we show the mechanism of $a_0(980)$, which is dynamically generated by the $S$-wave pseudoscalar-pseudoscalar interaction in Sec.~\ref{sec2b}. Finally, we will give the theoretical formalism for the double differential width of the process $\Lambda_c^+ \to \pi^+\eta n$ in Sec.~\ref{sec2c}.

\subsection{The $N(1535)$ role in $\Lambda_c^+ \to \pi^+\eta n$ reaction}\label{sec2a}

\begin{figure}[tbhp]
	\centering
	\includegraphics[scale=0.6]{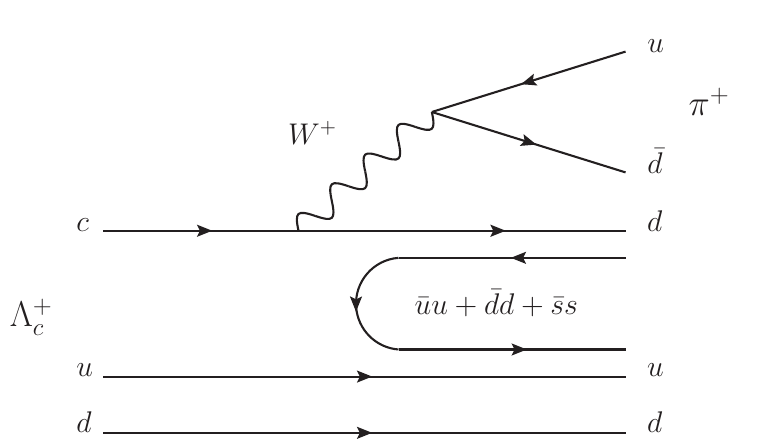}
	%\vspace{-0.7cm}
	\caption{Quark level diagram for the process $\Lambda_c^+ \to  \pi^{+}d(\bar{u}u+\bar{d}d+\bar{s}s)ud$ via the $W^+$ external emission.}
	\label{fig:hadronFSI}
\end{figure}

Taking into account that $N(1535)$ could be generated via meson-baryon interaction, as done in Refs.~\cite{Xie:2017erh,Li:2024rqb,Lyu:2023aqn,Li:2025exm}, we first show the dominant $ W^{+}$ external emission mechanism of the decay $\Lambda_c^+ \to \pi^+\eta n$ in Fig.~\ref{fig:hadronFSI}, the $c$ quark from the initial $\Lambda_c^+$ weakly decays into a ${W}^+$ boson and a $d$ quark, then the ${W^{+}}$ boson decays into $u\bar{d}$ quark pair. The $u\bar{d}$ quark pair from the ${W}^+$ boson will hadronize into $\pi^{+}$, while the $d$ quark and the $ud$ quark pair of the initial $\Lambda_c^+$ will hadronize with the antiquark-quark pair $\bar{q}q=\bar{u}u+\bar{d}d+\bar{s}s$, which is created from the vacuum with the quantum numbers $J^{PC}=0^{++}$, as follows,
\begin{align}
	\Lambda_{c}^+&= \frac{1}{\sqrt{2}}c(ud-du) \nonumber\\
	&\Rightarrow \frac{1}{\sqrt{2}}W^{+}d\left(ud-du\right) \nonumber\\
	&= \frac{1}{\sqrt{2}}u\bar{d}d\left(\bar{u}u+\bar{d}d+\bar{s}s\right)\left(ud-du\right) \nonumber\\
	&= \frac{1}{\sqrt{2}}\pi^{+}d\left(\bar{u}u+\bar{d}d+\bar{s}s\right)\left(ud-du\right) \nonumber\\
	&= \frac{1}{\sqrt{2}}\pi^{+}\sum_{i}M_{2i}q_{i}(ud-du),
\end{align}
where $i$=1, 2, 3 stand for the $u$, $d$ and $s$ quark, respectively, and $M$ is the SU(3) matrix of the octet of pseudoscalar mesons~\cite{Zhang:2024myn},
\begin{eqnarray}
	& M & = \left(\begin{array}{cccc}u\bar{u}&u\bar{d}&u\bar{s}\\
		d\bar{u}&d\bar{d}&d\bar{s}\\
		s\bar{u}&s\bar{d}&s\bar{s}
		\end{array}
		\right) \nonumber\\
	&&= \left(\begin{matrix} \frac{\eta}{\sqrt{3}}+ \frac{{\pi}^0}{\sqrt{2}}+ \frac{{\eta^\prime}}{\sqrt{6}} & \pi^+ & K^+  \\
		\pi^-  &   \frac{\eta}{\sqrt{3}}- \frac{{\pi}^0}{\sqrt{2}}+ \frac{{\eta^\prime}}{\sqrt{6}}  &  K^0 \\
		K^-  &  \bar{K}^{0}   &    -\frac{\eta}{\sqrt{3}}+ \frac{{\sqrt{6}\eta^\prime}}{3}
	\end{matrix}
	\right),
\end{eqnarray} 
where we utilize the $\eta-\eta^{\prime}$ mixing following Refs.~\cite{Bramon:1992kr,Lyu:2023ppb}, and the $\eta^{\prime}n$ channel is ignored since the $\eta^{\prime}$ has a large mass.

Next, we could write all the possible final  meson-baryon components after the hadronization,
\begin{align}\label{lambda-dianhe}
	\Lambda_{c}^{+} =  \pi^{+}\left(\pi^{-}p-\frac{1}{\sqrt{2}}\pi^{0}n+\frac{1}{\sqrt{3}}\eta n-\sqrt{\frac{2}{3}}K^{0}\Lambda\right),
\end{align}
where we adopt the flavor-wave functions of the baryon as~\cite {Pavao:2017cpt,Miyahara:2016yyh}, 
\begin{gather}
		p=	\frac{u(ud-du)}{\sqrt{2}}, ~~~		n=\frac{d(ud-du)}{\sqrt{2}}, \\
		\Lambda=\frac{u(ds-sd)+d(su-us)-2s(ud-du)}{2\sqrt{3}}.
\end{gather} 

Considering the isospin multiplets $(-\pi^+,\pi^0,\pi^-)$ and $(p,n)$, we could write,
\begin{align}
	\pi^-p&=|1,-1\rangle\left|\frac12,\frac12\right\rangle \nonumber\\
	&=\sqrt{\frac13}\left|\frac32,-\frac12\right\rangle-\sqrt{\frac23}\left|\frac12,-\frac12\right\rangle,
\end{align}
\begin{align}
	\pi^0n&=|1,0\rangle\left|\frac12,-\frac12\right\rangle \nonumber\\
	&=\sqrt{\frac23}\left|\frac32,-\frac12\right\rangle+\sqrt{\frac13}\left|\frac12,-\frac12\right\rangle,
\end{align}
\begin{align}
	\pi^-p-\frac{1}{\sqrt2}\pi^0n=&\left(\sqrt{\frac13}-\sqrt{\frac12}\cdot\sqrt{\frac23}\right)|\pi N\rangle^{I=\frac32} \nonumber\\
	&+\left(-\sqrt{\frac23}-\sqrt{\frac12}\cdot\sqrt{\frac13}\right)|\pi N\rangle^{I=\frac12} \nonumber\\
	=&-\sqrt{\frac32}\left|\pi N\right\rangle^{I=\frac12}.
\end{align}

In the isospin basis, we can obtain all potential components of the final states,
\begin{align}
	\Lambda_{c}^{+} = \pi^{+}\left(-\sqrt{\frac{3}{2}}\pi N+\sqrt{\frac{1}{3}}\eta N-\sqrt{\frac{2}{3}}K^{0}\Lambda\right).
\end{align}
\begin{figure}[htbp]
		 \centering		
		\subfigure[]{
			\centering
			\includegraphics[scale=0.6]{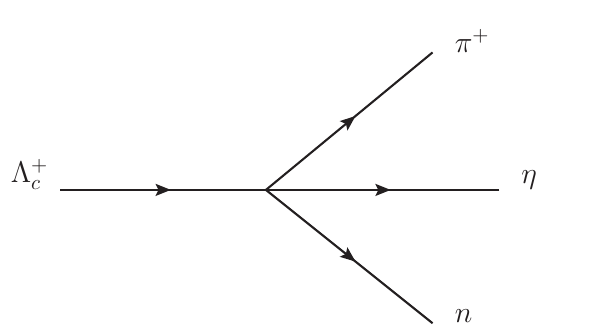}
		}
		\subfigure[]{
		    \centering
			\includegraphics[scale=0.6]{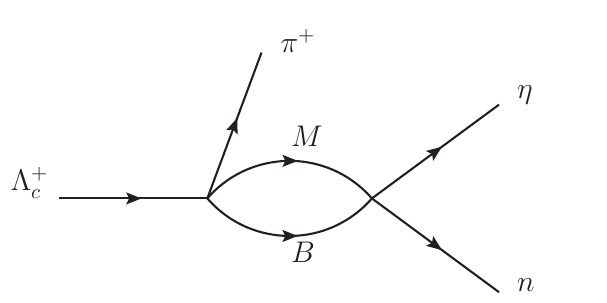}
		}
		\subfigure[]{
		    \centering
			\includegraphics[scale=0.6]{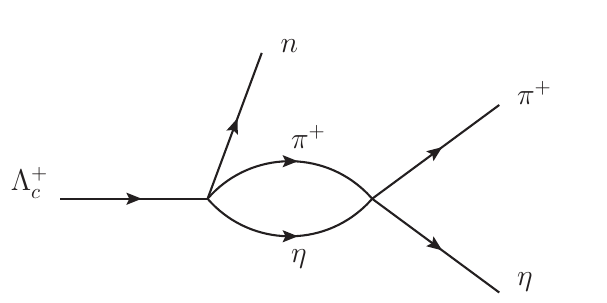}
		}
		\caption{The mechanisms of the process $\Lambda_c^+ \to  \pi^+\eta n$. (a) Tree level, (b) the $S$-wave pseudosalar meson-octet baryon interaction, (c) the $S$-wave $\pi^+\eta$ final state interaction.}\label{fig:tree and quan}
\end{figure}
	
Then, the decay $\Lambda_c^+ \to  \pi^+\eta n$ could happen through the $S$-wave pseudoscalar meson-octet meson interaction, which will dynamically generate the $N(1535)$ state, as show in Fig.~\ref{fig:tree and quan}(b), and the decay amplitude could be written as,
\begin{equation}\label{M-N1535}
	\begin{aligned}
		\mathcal{M}^{N(1535)} &= \sum_{i}h_{i}G_{i}t_{i\rightarrow\eta N},  
	\end{aligned}
\end{equation}
where  $i=1, 2, 3$ correspond to the $\pi N$, $\eta N$, and $K^0\Lambda$ channel, respectively, and
\begin{equation}
	h_{\pi N} =-\sqrt{\frac{3}{2}}, \quad h_{\eta N}=\sqrt{\frac{1}{3}}, \quad h_{K^0\Lambda}=-\sqrt{\frac{2}{3}}.
\end{equation}

The $G_i$ in Eq.~(\ref{M-N1535}) is the loop function of the meson-baryon system~\cite{Inoue:2001ip},
\begin{equation}\label{loop function}
G_i=i\int\frac{d^4q}{(2\pi)^4}\frac{2M_i}{(P-q)^2-M_i^2+i\epsilon}\frac{1}{q^2-m_i^2+i\epsilon},
\end{equation}
where $M_{i}$ and $m_{i}$ are the masses of baryon and meson of $i$-th coupled channel, respectively. $P$ is the four-momentum of the meson-baryon system, and $q$ is the four-momentum of meson in the center-of-mass frame. For the meson-baryon loop function, we take the cut-off method~\cite{Guo:2005wp}, and the loop function could be written as,  
\begin{equation}
\begin{aligned}
		G(s)=&\frac{1}{16\pi^{2}s}\left\{\sigma\left(\arctan\frac{s+\Delta}{\sigma\lambda_{1}}+\arctan\frac{s-\Delta}{\sigma\lambda_{2}}\right)\right. \\
		&\left.-\left[\left(s+\Delta\right)\ln\frac{q_{max}\left(1+\lambda_{1}\right)}{m_{1}}+\left(s-\Delta\right)\ln\frac{q_{max}\left(1+\lambda_{2}\right)}{m_{2}}\right]\right\},
\end{aligned}
\end{equation}
where
\begin{equation}
\sigma=\left[-\left(s-(M_{i}+m_{i})^{2}\right)\left(s-(M_{i}-m_{i})^{2}\right)\right]^{1/2},
\end{equation}
\begin{equation}
\Delta=M_{i}^{2}-m_{i}^{2},
\end{equation}
\begin{equation}
\lambda_1=\sqrt{1+\frac{M_{i}^{2}}{q_{\max}^{2}}},~~~~\lambda_2=\sqrt{1+\frac{m_{i}^{2}}{q_{\max}^{2}}}.
\end{equation}
In order to dynamically generate $N(1535)$ state, we adopt the cut-off parameter $q_\mathrm{max}=1150$ MeV as Ref.~\cite{Li:2024rqb}. 

The $t_{i\to \eta N}$ in Eq.~(\ref{M-N1535}) are the transition amplitude of the coupled channels, and $t_{i\to \eta N}$ could acquire through the Bethe-Salpeter equation,
\begin{equation}\label{BS}
	T=[1-VG]^{-1}V.
\end{equation}
In our calculation, we take into account four coupled channels $\pi N$, $\eta N$, $K\Lambda$, and $K\Sigma$. The transition potential $V_{ij}$ is obtained from Ref.~\cite{Wang:2015pcn},
\begin{equation}
	\begin{aligned}
		V_{ij}&= -C_{ij}\frac{1}{4f_{i}f_{j}}\left(2\sqrt{s}-M_{i}-M_{j}\right)\\&\times\sqrt{\frac{E_{i}+M_{i}}{2M_{i}}}\sqrt{\frac{E_{j}+M_{j}}{2M_{j}}},
	\end{aligned}
\end{equation}
where $M_{i}$ and $E_{i}$ correspond to the mass and energy of baryon in $i$ channel, 
\begin{equation}
	E_{i}=\frac{s+M_{i}^2-m_{i}^2}{2\sqrt{s}}.
\end{equation}
%where $s=M_{\eta n}^2$. %is the invariant mass squared of the meson-baryon system.

\begin{table}[htbp]
	\begin{center}	
	\caption{The $S$-wave meson-baryon scattering coefficients~\cite{Li:2024rqb}.} 	\label{tab:Cij}
	\begin{tabular}{ccccc}%五个c表示有4列
		\hline\hline  
		            \qquad\quad& $\pi N$  \qquad\quad& $\eta n$  \qquad\quad& $K\Lambda$  \qquad\quad& $K\Sigma$ \\ \hline
		$\pi N$     \qquad\quad&2         \qquad\quad&0          \qquad\quad&3/2          \qquad\quad&-1/2      \\
		$\eta n$    \qquad\quad&          \qquad\quad&0          \qquad\quad&-3/2         \qquad\quad&-3/2       \\
		$K\Lambda$  \qquad\quad&          \qquad\quad&           \qquad\quad&0            \qquad\quad&0          \\
		$K\Sigma$   \qquad\quad&          \qquad\quad&           \qquad\quad&             \qquad\quad&2          \\
		\hline\hline
	\end{tabular}
	\end{center}
\end{table}

The values of coefficients $C_{ij}$ are presented in Table~\ref{tab:Cij}. The $f_{i}$ are the meson decay constants of $i$ channel, which are given as,
\begin{equation}
	f_\pi=93~\mathrm{MeV},\quad f_K=1.22f_\pi,\quad f_\eta=1.3f_\pi.
\end{equation}

\subsection{The $a_0(980)$ role in $\Lambda_c^+ \to  \pi^+\eta n$ reaction}\label{sec2b}

In this subsection, based on the discussions of Fig.~\ref{fig:hadronFSI} and Eq.~(\ref{lambda-dianhe}), we can also consider $\pi^{+}\eta$ final state interaction, which will dynamically generate the scalar state $a_0(980)$, as shown in Fig.~\ref{fig:tree and quan}(c), and the decay amplitude could be expressed as,
\begin{equation}\label{M-pieta1}
	\begin{aligned}
		\mathcal{M}_{a} = h_{\pi^{+}\eta}G_{\pi^{+}\eta}t_{\pi^{+}\eta\rightarrow\pi^{+}\eta},  
	\end{aligned}
\end{equation}
where $h_{\pi^{+}\eta}$=$\sqrt{1/3}$ from Eq.~(\ref{lambda-dianhe}), and the transition amplitude $t_{\pi^{+}\eta\rightarrow\pi^{+}\eta}$ could be obtained by solving the Bethe-Salpeter equation as Eq.~(\ref{BS}). Here, $V$ represents a $2\times2$ matrix for the interaction kernal with two coupling channels $K^+\bar{K}^0$ and $\pi^+\eta$, and the matrix elements are as follows~\cite{Lin:2021isc,Duan:2024led},
\begin{align}
		&V_{K^{+}\bar{K}^{0}\to K^{+}\bar{K}^{0}}=-\frac{s}{4f^{2}}, \nonumber\\
		&V_{K^{+}\bar{K}^{0}\to\pi^{+}\eta}=-\frac{3s-2m_K^2-m_\eta^2}{3\sqrt{3}f^2},\nonumber\\
		&V_{\pi^+\eta\to\pi^+\eta}=-\frac{2m_{\pi}^2}{3f^2},
\end{align}
where $s=M^2_{\text{inv}}(\pi^+\eta)$, and $f=f_{\pi}=93$~MeV which is the pion decay constant. $m_\pi$  and $m_K$ represent the averaged masses of the pion and kaon.

For meson-meson system, the loop function $G$ is given by
the cut-off method with the cut-off parameter $q_{\mathrm{max}}=600$ MeV~\cite{Xie:2014tma}, and it can be expressed as
\begin{equation}\label{Eq:G-meson-meson}
\begin{aligned}
	G &= i\int\frac{d^{4}q}{(2\pi)^{4}}\frac{1}{(P-q)^{2}-m_{1}^{2}+i\epsilon}\frac{1}{q^{2}-m_{2}^{2}+i\epsilon}  \\
	&=\int_{0}^{q_{\mathrm{max}}}\frac{|\vec{q}|^{2}d|\vec{q}|}{(2\pi)^{2}}\frac{\omega_{1}+\omega_{2}}{\omega_{1}\omega_{2}{[s-(\omega_{1}+\omega_{2})^{2}+i\epsilon]}},
\end{aligned}
\end{equation}
where the meson energies $\omega_{i}$=$\sqrt{\left|\vec{q}\right|^{2}+m_{i}^{2}}$.  %Based on , we using the three-momentum cut-off method to solve the singular integral of $G$ 

\begin{figure}[tbhp]
 	\centering
 	\includegraphics[scale=0.6]{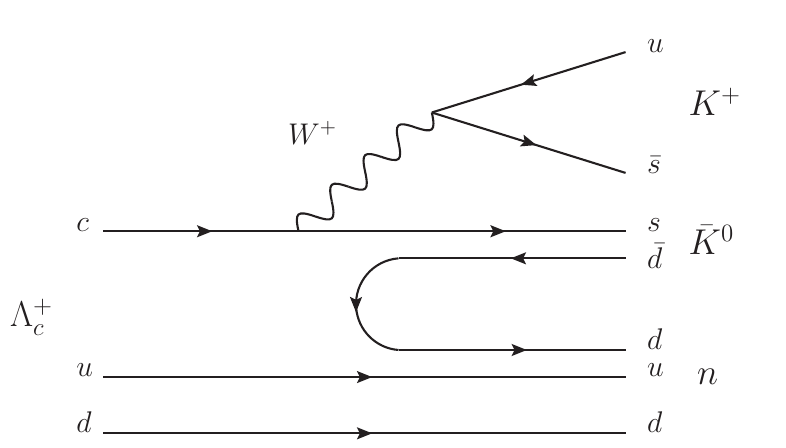}
 	%\vspace{-0.7cm}
 	\caption{Quark level diagram for the process $\Lambda_c^+ \to  K^+\bar{K}^{0}n$ via the $W^+$ external emission.}
 	\label{fig:level}
 \end{figure}

%Furthermore, as shown in Fig.~\ref{fig:level-pieta}, the $c$ quark from the initial $\Lambda_c^+$ weakly decays into a $s$ quark and a $W^+$ boson, following that the $W^+$ boson subsequently decays into a $u\bar{s}$ quark pair, which will hadronize into a $K^+$ meson. The $s$ quark from the $\Lambda_c^+$ decay and the $ud$ quark pairs of the initial $\Lambda_c^+$, together with the quark pair $\bar{d}d$ created from the vacuum with the quantum numbers $J^{PC}=0^{++}$, hadronize into $\bar{K}^0n$, then endures the re-scattering $K^{+}\bar{K}^{0} \to \pi^{+}\eta$. This approach can also generate the final state $\pi^{+}\eta n$.
In addition to the mechanism in Fig.~\ref{fig:hadronFSI} that we have considered, the decay $\Lambda_c^+ \to  \pi^+\eta n$ could happen through the re-scattering $\Lambda_c^+\to K^{+}\bar{K}^{0}n \to \pi^{+}\eta n$, as showed in Fig.~\ref{fig:level}, i.e. the $c$ quark from the initial $\Lambda_c^+$ weakly decays into an $s$ quark and a $W^+$ boson, following that the $W^+$ boson subsequently decays into a $u\bar{s}$ quark pair, which will hadronize into a $K^+$ meson. The $s$ quark from the $\Lambda_c^+$ decay and the $ud$ quark pairs of the initial $\Lambda_c^+$, together with the quark pair $\bar{d}d$ created from the vacuum with the quantum numbers $J^{PC}=0^{++}$, hadronize into $\bar{K}^0n$, then endures the re-scattering $K^{+}\bar{K}^{0} \to \pi^{+}\eta$. One can see that now $K^{+}\bar{K}^{0}$ can interact and give $\pi^{+}\eta$.
 \begin{figure}[tbhp]
 	\centering
 	\includegraphics[scale=0.6]{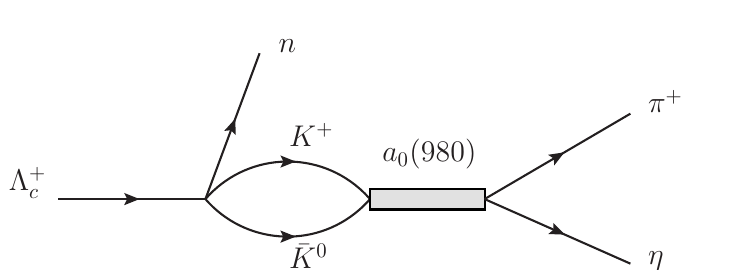}
 	%\vspace{-0.7cm}
 	\caption{The mechanism of the process $\Lambda_c^+ \to \pi^+\eta n$ via the re-scattering $K^{+}\bar{K}^{0} \to \pi^{+}\eta$.}
 	\label{fig:pieta-quan2}
 \end{figure}

%Then, the decay $\Lambda_c^+ \to \pi^+\eta n$ could happen through the $S$-wave pseudoscalar-pseudoscalar interaction of Fig.~\ref{fig:pieta-quan2}, and the amplitude can be given by,
Meanwhile, one can find that the mechanism in Fig.~\ref{fig:level} is in proportion to the Cabbibo-Kobayashi-Maskawa (CKM) matrix elements $V_{cs}V_{us}$, and the mechanism in Fig.~\ref{fig:hadronFSI} is in proportion to the CKM matrix elements $V_{cd}V_{ud}$. Considering $|V_{cs}|\approx |V_{ud}|$ and $|V_{cd}|\approx |V_{us}|$, we deem that both the mechanisms should be of the same order of magnitude. Subsequently, we easily account for this decay mode of Fig.~\ref{fig:pieta-quan2} and the amplitude can be given by, 
\begin{equation}\label{M-pieta2}
	\begin{aligned}
		\mathcal{M}_{b} = h_{K^{+}\bar{K}^{0}n}G_{K^{+}\bar{K}^{0}}t_{K^{+}\bar{K}^{0}\rightarrow\pi^{+}\eta},
	\end{aligned}
\end{equation} 
with $h_{K^{+}\bar{K}^{0}n}=1$, where we calculate the meson-meson loop function $G$ using a same way as Eq.~(\ref{Eq:G-meson-meson}).

Therefore, the total amplitude for the $a_{0}(980)$ state is
\begin{equation}\label{Eq:M-a0}
\mathcal{M}^{a_0(980)}=\mathcal{M}_{a}+\mathcal{M}_{b}.
\end{equation}

\begin{figure}[htbp]	
	\subfigure[]{
		\centering
		\includegraphics[scale=0.6]{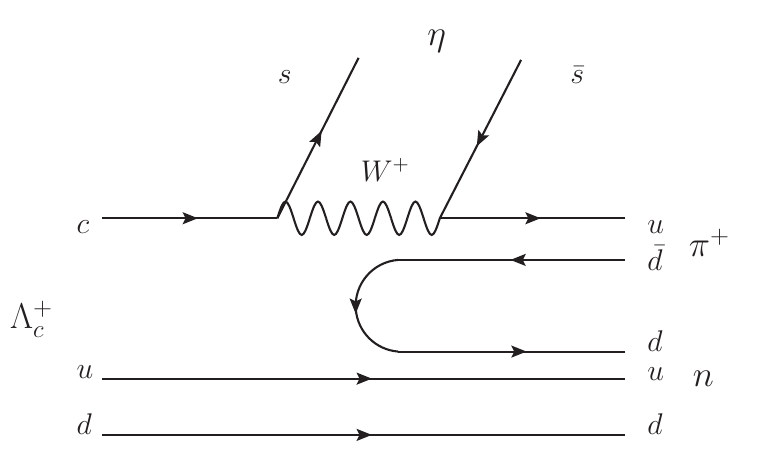}	
		\label{fig:level-last1}
	}
	\subfigure[]{
		\centering
		\includegraphics[scale=0.6]{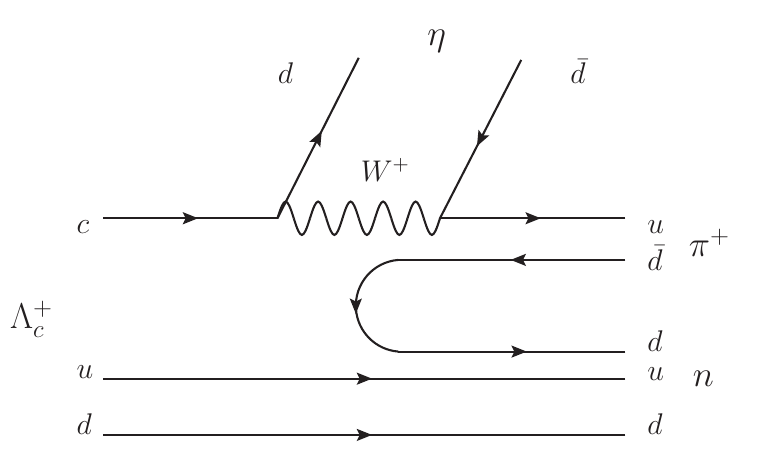}	
		\label{fig:level-last2}
	}
	\caption{Quark level diagrams for the process $\Lambda_c^+ \to  \pi^+\eta n$ via the $W^+$ internal emission.}
	\label{fig:ie}
\end{figure}

It should be noted that the $\Lambda_c^+ \to  \pi^+\eta n$ process could happen through $W^{+}$ internal emission in Fig.~\ref{fig:ie}. However, in our calculation, we have ignored these mechanisms, since the internal emission has $1/N_{c}$ suppression relative to the external excitation~\cite{Lyu:2023ppb,Li:2023nsw,Zhang:2022xpf,Duan:2024led}, and it is expected to give a smaller contribution. On the other hand, many theories propose that $N(1535)$ state has an $\bar{s}s$ component, while $dud$ does not have an $\bar{s}s$ component. Meanwhile, considering that there is no experimental data available for the process $\Lambda_c^+ \to \pi^+\eta n$, it will increase the number of the free parameters if we include mechanism of $W^{+}$ internal emission. Therefore, we will ignore the mechanism of $W^{+}$ internal emission at present. 
%In Fig.~\ref{fig:level-last1}, the hadronization of $u$ qurak and the $ud$ quark pair from the initial $\Lambda_c^+$, along with the $\bar{d}d$ pair generated from the vacuum as follows, 
%\begin{equation}
%\begin{aligned}
%		\Lambda_{c} & \Rightarrow s\bar{s}u\left(\bar{d}d\right)\frac{1}{\sqrt{2}}\left(ud-du\right) \\
%		& =\left(M_{33}\times M_{12}\right)\frac{1}{\sqrt{2}}\left(ud-du\right) \\
%		& =\left(M_{33}\times M_{12}\right)n \\
%		& =-\frac{1}{\sqrt{3}}\eta\pi^{+}n,
%\end{aligned}
%\end{equation}
%where the c quark weakly decay $W^{+}$ bason through the internal emission, and the internal emission has $1/N_{c}$  suppression relative to the external excitation shown in Fig.~\ref{fig:level-pieta}~\cite{Lyu:2023ppb}. The interaction between $\pi^{+}$ and $n$ is the dominant aspect, almost all nucleon excited states can decay into $\pi^{+}n$ channel. Meanwhile, since there is no experimental data available for the process $\Lambda_c^+ \to \pi^+\eta n$, it is impossible for us to take all of them into consideration. In addition, some theories propose that $N(1535)$ has a $\bar{s}s$ component, while $dud$ does not have an $\bar{s}s$ component. Therefore, we will not consider the interaction of $\pi^{+}n$ at present.

\subsection{Invariant mass distributions}\label{sec2c}

Using the formalisms mentioned above, we can detail the total decay amplitude of process $\Lambda_c^+ \to  \pi^+\eta n$ as follows, 
\begin{eqnarray}\label{M-total}
 \mathcal{M} =\mathcal{M}^{\text{Tree}}+\mathcal{M}^{N(1535)}+\mathcal{M}^{a_0(980)},
\end{eqnarray}
where 
\begin{equation}\label{Eq:Tree}
    \mathcal{M}^{\text{Tree}}=h_{\pi^{+}\eta},
\end{equation}
and the double differential width of the process $\Lambda_c^+ \to  \pi^+\eta n$ is
\begin{eqnarray}
    \frac{d^{2}\Gamma}{dM_{\eta n} dM_{\pi^{+}\eta}}&=\dfrac{1}{(2\pi)^{3}}\dfrac{M_{n}}{2M_{\Lambda_c^+}^2}|\mathcal{M}|^{2}M_{\eta n} M_{\pi^{+}\eta}, \label {eq:dgammadm12dm23} 
\end{eqnarray}
where the square of the modulus of the total amplitude represents, 
\begin{eqnarray}\label{Eq:M-total}
%|\mathcal{M}|^2=|\mathcal{M}^{tree}+\mathcal{M}^{N(1535)}+\mathcal{M}^{a_0(980)}e^{i\phi}|^2.
|\mathcal{M}|^2=|\mathcal{M}^{\text{Tree}}+\mathcal{M}^{N(1535)}+\mathcal{M}^{a_0(980)}|^2.
\end{eqnarray}
%and there is a phase $\phi$ between the three terms. 

Given a specific value of the invariant mass $M_{12}$, the corresponding range for the invariant mass $M_{23}$ is determined according to the Review of Particle Physics (RPP)~\cite{ParticleDataGroup:2024cfk}
\begin{align}
	&\left(m_{23}^2\right)_{\min}=\left(E_2^*+E_3^*\right)^2-\left(\sqrt{E_2^{* 2}-m_2^2}+\sqrt{E_3^{* 2}-m_3^2}\right)^2, \nonumber\\
	&\left(m_{23}^2\right)_{\max}=\left(E_2^*+E_3^*\right)^2-\left(\sqrt{E_2^{* 2}-m_2^2}-\sqrt{E_3^{* 2}-m_3^2}\right)^2, \label{eq:limit}
\end{align}
where $E_2^*$ and $E_3^*$ are the energies of particles 2 and 3 in the $M_{12}$ rest frame, respectively,
\begin{align}
	E_{2}^{*}&=\frac{M_{12}^{2}-m_{1}^{2}+m_{2}^{2}}{2M_{12}}, \nonumber\\
	E_{3}^{*}&=\frac{M_{\Lambda_c^{+}}^{2}-M_{12}^{2}-m_{3}^{2}}{2M_{12}}.
\end{align}
where $m_1$, $m_2$, and $m_3$ denote the masses of particles 1, 2, and 3,  respectively. The masses and widths of the particles are sourced from the RPP~\cite{ParticleDataGroup:2024cfk}.

It is notable that there could be contributions from other excited nucleons. For instance, there is a state $N(1520)$ with $J^P=3/2^-$ close to the $N(1535)$, which should be suppressed since this state couples to the $\eta n$ or $\pi n$ in $D$-wave. Furthermore, it will increase the number of the free parameters if we take into account more intermediate resonances, and one can consider them when the precise measurements of this process are available in the future.

%%%%%%%%%%%%%%%%%%%%%%
\section{Results and Discussion} \label{sec:Results}

\begin{figure}[htbp]	
	\subfigure{
		\centering
		\includegraphics[scale=0.65]{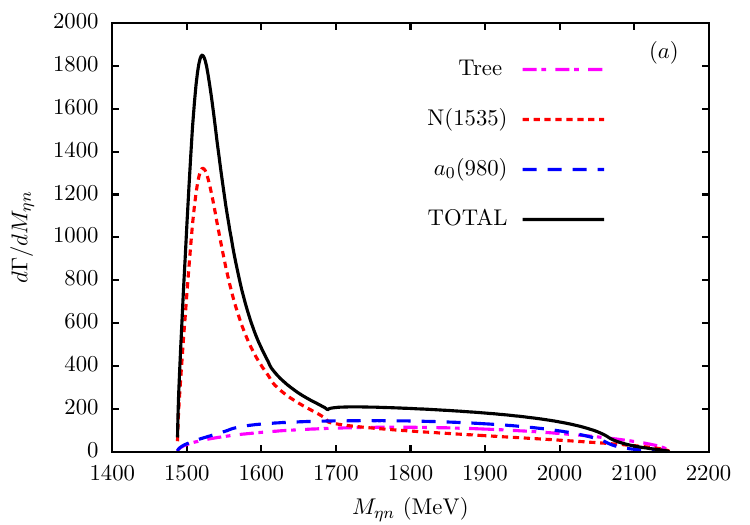}	
		%\label{fig:minv1}
	}
	\subfigure{
		\centering
		\includegraphics[scale=0.65]{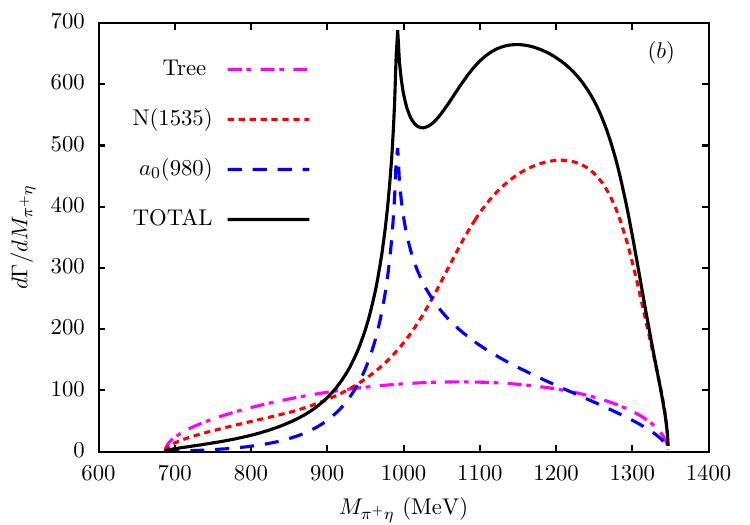}	
		%\label{fig:minv2}
	}
	\caption{The $\eta n$ (a) and $\pi^+\eta$ (b) invariant mass distributions of the process $\Lambda_c^+ \to \pi^+\eta n$ decay.}
	\label{fig:minv}
\end{figure}
\begin{figure}[htbp]
	\centering		
	\subfigure{
		\centering
		\includegraphics[scale=0.65]{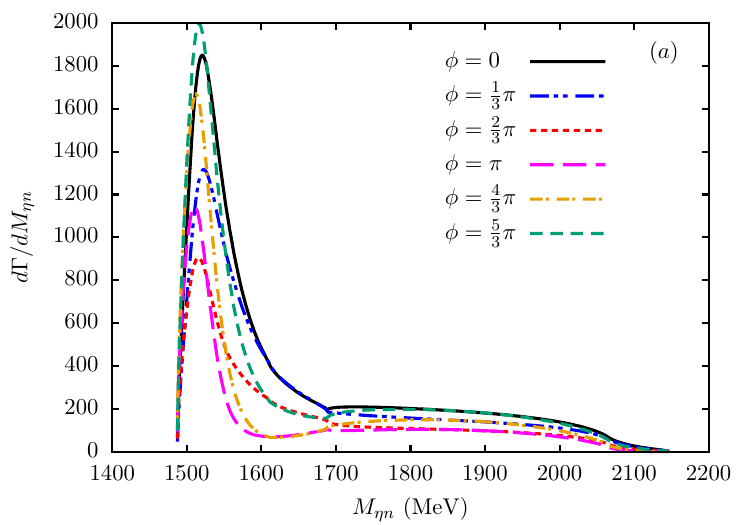}	
		%\label{fig:minv1-phi}
	}
	\subfigure{
		\centering
		\includegraphics[scale=0.65]{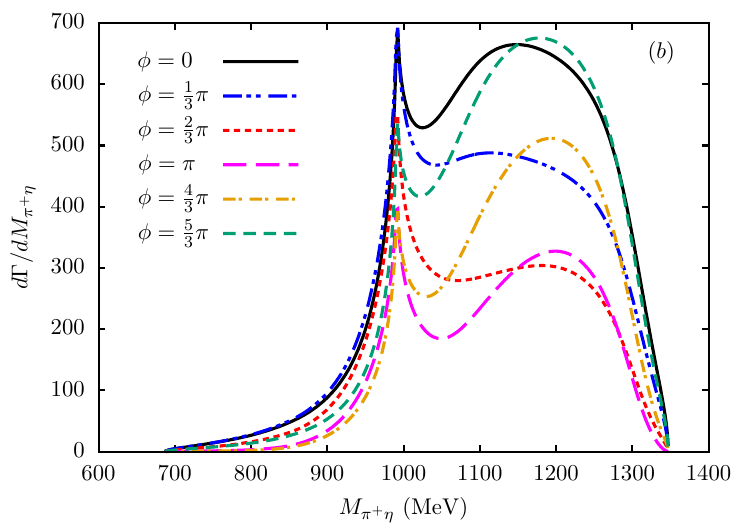}	
		%\label{fig:minv2-phi}
	}
	\caption{The $\eta n$ (a) and $\pi^+\eta$ (b) invariant mass distributions of the process $\Lambda_c^+ \to \pi^+\eta n$ decay with the interference phase $\phi=0, \pi/3, 2\pi/3, \pi, 4\pi/3$, and $5\pi/3$, respectively.}
	\label{fig:minv-phi}
\end{figure}

\begin{figure}[htbp]
%	\centering		
%	\subfigure{
%		\centering
		\includegraphics[scale=0.80]{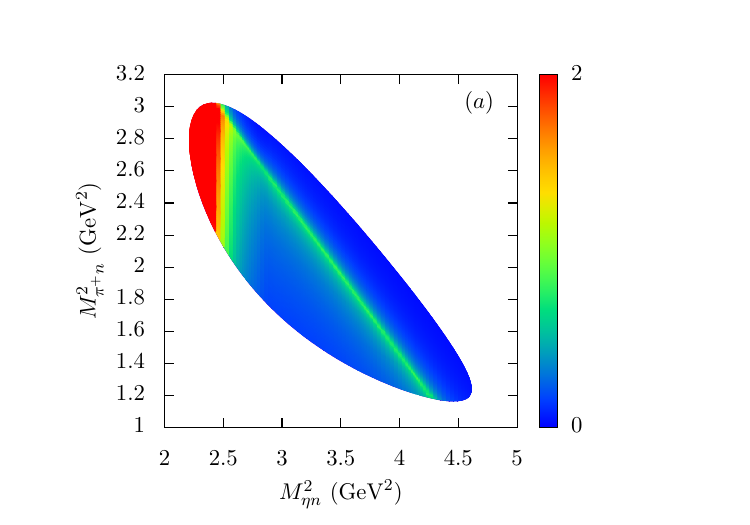}	
		%\label{fig:dalitz1}
%	}%\label{fig:dalitz1}
%	\subfigure{
%		\centering
		\includegraphics[scale=0.80]{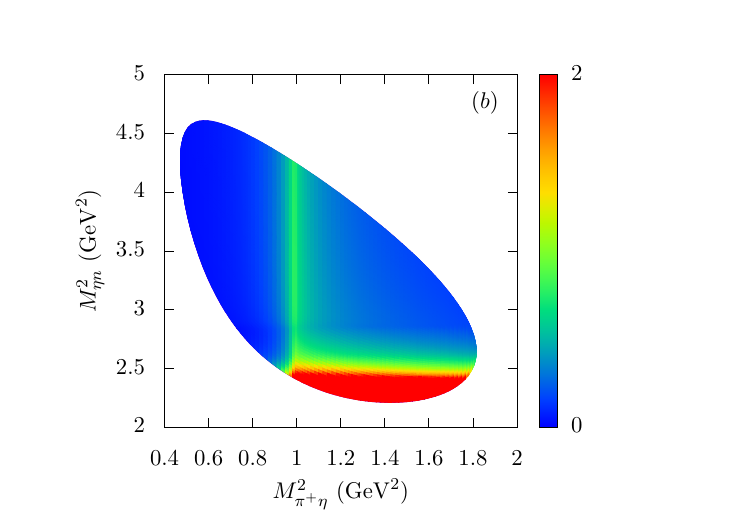}	
		%\label{fig:dalitz2}
%	}%\label{fig:dalitz2}	
%	\subfigure{
%		\centering
		\includegraphics[scale=0.80]{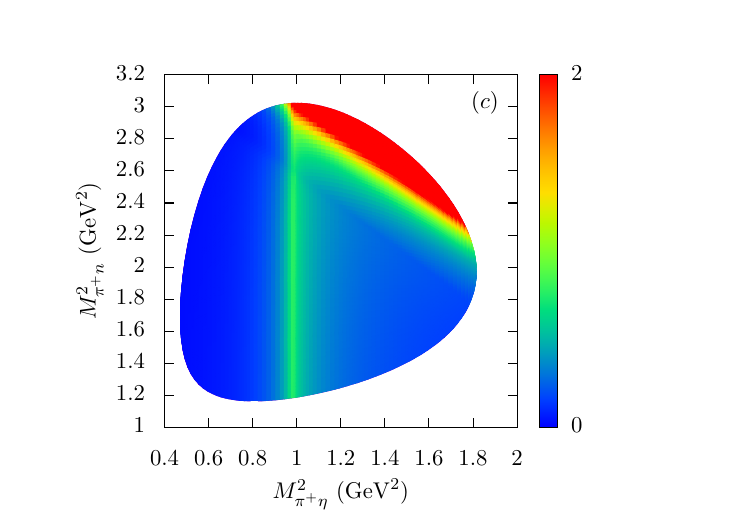}
	%	\label{fig:dalitz3}    
%	}%\label{fig:dalitz3}
	\caption{The Dalitz plots for the process $\Lambda_c^+ \to \pi^+\eta n$.  (a) $M_{\eta n}^2$ vs. $M_{\pi^+n}^2$; (b) $M_{\pi^+\eta}^2$ vs. $M_{\eta n}^2$; (c) $M_{\pi^+\eta}^2$ vs. $M_{\pi^+n}^2$.}
	\label{fig:dalitz}
\end{figure}

As we discussed above, we can get $d\Gamma/dM_{12}$ by integrating $d^2\Gamma/(dM_{12}dM_{23})$ over $M_{23}$ with the limits of Eq.~(\ref{eq:limit}). It is worth mentioning that, in our model, there is no free parameter. Firstly, we show our results of the $\eta n$ and $\pi^{+}\eta$ invariant mass distribution in Figs.~\ref{fig:minv}(a) and \ref{fig:minv}(b), respectively. The magenta-dot-dashed curves show the contribution from the tree level of Eq.~(\ref{Eq:Tree}). The red-dotted curves show the contribution from the intermediate $N(1535)$ state of Eq.~(\ref{M-N1535}). The blue-dashed curves show the contribution from the $a_{0}(980)$ state of Eq.~(\ref{Eq:M-a0}). The black-solid curves show the total contribution from Eq.~(\ref{Eq:M-total}). One can find a significant near-threshold enhancement structure in $\eta n$ invariant mass distribution, which is due to the nucleon resonance $N(1535)$. On the other hand, one also can find a clear cusp structure around 980~MeV in $\pi^+\eta$ invariant mass distribution, which could be associated with $a_{0}(980)$. Meanwhile, we can see a bump structure standing for the region of 1100$\sim$1300~MeV in $\pi^+\eta$ invariant mass distribution resulting from the reflection of the $N(1535)$ state.

%We first present our results of $\eta n$ and $\pi^{+}\eta$ invariant mass distribution in Figs.~\ref{fig:minv1} and \ref{fig:minv2}, where the magenta-dot-dashed curve shows the contribution of tree diagram, the red-dotted curves and the blue-dashed curve stand for the contribution of $N(1535)$ state and $a_{0}(980)$ state respectively, and the last black-solid curve indicates the total contribution. In Fig.~\ref{fig:minv1}, there is a significant peak structure of $N(1535)$. Meanwhile, we can find a clear cusp structure around 980 MeV in the $\pi^{+}\eta$ invariant mass distribution, which is associated to the $a_{0}(980)$. 

In addition, it should be noted that there may be phase interference between different contributions of Eq.~(\ref{Eq:M-total}). Thus, we multiply the amplitude $\mathcal{M}^{a_0(980)}$ in Eq.~(\ref{Eq:M-total}) by a phase factor $e^{i\phi}$, and calculate the invariant mass distributions with $\phi=0$, $\pi/3$, ${2\pi}/{3}$, $\pi$, ${4\pi}/{3}$, and ${5\pi}/{3}$, as presented in Figs.~\ref{fig:minv-phi}(a) and \ref{fig:minv-phi}(b). Although the lineshape is distorted by interference with different phase angles $\phi$, one can always find the near-threshold enhancement structure of the $N(1535)$ state in $\eta n$ mass distribution and the cusp structure of the $a_{0}(980)$ in $\pi^+\eta$ mass distribution.

%In our work, we only take one free parameter is the phase angle $\phi$ of Eq.~(\ref{eq:total-M}) with  $\phi=0, \pi/3, 2\pi/3, \pi, 4\pi/3$, and $5\pi/3$. We perform our results in Figs.~\ref{fig:minv1-phi} and \ref{fig:minv2-phi}, one can find that, with different value of the phase $\phi$, there is still a peak or cusp structure in $\eta n$ and $\pi^{+}\eta$ invariant mass distribution.

Next, we present the Dalitz plots of ``$M_{\eta n}^2$" vs. ``$M_{\pi^+n}^2$", ``$M_{\pi^+\eta}^2$" vs. ``$M_{\eta n}^2$" and ``$M_{\pi^+\eta}^2$" vs. ``$M_{\pi^+n}^2$" for the process $\Lambda_c^+ \to \pi^+\eta n$ in Figs.~\ref{fig:dalitz}(a), ~\ref{fig:dalitz}(b), and ~\ref{fig:dalitz}(c), respectively. One can clearly find the signals of the resonance $N(1535)$ and the scalar meson $a_0(980)$. 

%In Fig.~\ref{fig:dalitz}, we show the Dalitz plots for the process $\Lambda_c^+ \to \pi^+\eta n$. We can clearly see that there is a vertical red band near $M_{\eta n}=1535$ MeV in Figs.~\ref{fig:dalitz1} and \ref{fig:dalitz2}, which should be associated with the signal of the resonance $N(1535)$. In Figs.~\ref{fig:dalitz2} and \ref{fig:dalitz3}, we find a green band around $M_{\pi^+\eta}=980$ MeV, corresponding to the signal of the scalar $a_0(980)$.

\begin{figure}[tbhp]
	\centering
	\includegraphics[scale=0.65]{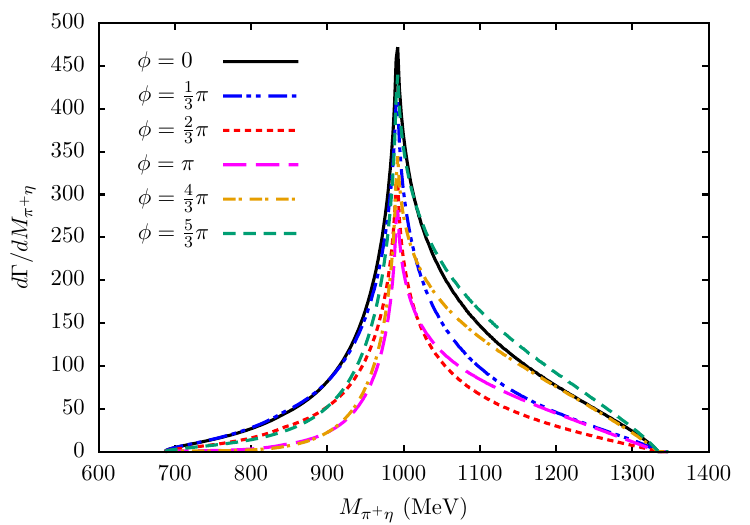}
	%\vspace{-0.7cm}
	\caption{The $\pi^{+}\eta$ invariant mass distribution of the process $\Lambda_c^+ \to \pi^+\eta n$ decay restricted to $M_{\eta n}\geq1700$ MeV.}
	\label{fig:cut-phi}
\end{figure}

Then, considering that the structure of $a_0(980)$ state is close to the signal of $N(1535)$ in $\pi^+\eta$ invariant mass distribution, in order to display the contribution of $a_0(980)$ more clearly, we take the cut of $M_{\eta n}\geq1700$~MeV to eliminate the contributions from the $N(1535)$ and show the results in Fig.~\ref{fig:cut-phi}. We can observe that under different phase angles, there will be a significant cusp in the $\pi^+\eta$ invariant mass distribution, which could be used to determine the strength of the $a_0(980)$.

%Besides, in Fig.~\ref{fig:minv2-phi}, there is a structure from 1050 MeV to 1250 MeV in $\pi^+\eta$ invariant mass distribution due to the reflection of the $N(1535)$ signal. In order to display the contribution of $a_0(980)$ more clearly, we made a cut by removing the part where the $M_{\eta n}$ is less than 1700 MeV. Then, we can obtain an invariant mass distribution with only the contribution of $a_0(980)$, as shown in Figure Fig.~\ref{fig:cut-phi}. We can observe that under different phase angles, there will be a significant cusp in the $\pi^+\eta$ invariant mass distribution. Additionally, We further showed the Dalitz plot after the cut in Fig.~\ref{fig:dali-cut}, with a more noticeable signal close to 980 MeV.

Further, we can integrate the invariant masses $M_{\eta n}$ and $M_{\pi^+\eta}$ over the whole invariant mass range for the signal of $a_0(980)$ and the total amplitude, and their ratio is given by,
\begin{equation}
    R=\dfrac{\mathcal{B}(\Lambda_c^+ \to a^+_0(980) n)}{\mathcal{B}(\Lambda_c^+ \to \pi^+\eta n)}=\dfrac{\int\frac{ d^{2}\Gamma_{a_0(980)}}{dM_{\eta n} dM_{\pi^{+}\eta}}}{\int\frac{ d^{2}\Gamma_{total}}{dM_{\eta n} dM_{\pi^{+}\eta}}}\approx 0.313,
\end{equation}
where $\Gamma_{a_0(980)}$ denotes that the width includes the contribution of the $a_0(980)$ only, i.e. the $\mathcal{M}=\mathcal{M}^{a_0(980)}$, and the $\Gamma_{total}$ denotes that the width includes all contributions, i.e. using the $\mathcal{M}$ of Eq.~(\ref{Eq:M-total}). The ratio is consistent with the predicted  $\mathcal{B}(\Lambda_c^+\to\Lambda a_0(980))/\mathcal{B}(\Lambda_c^+\to\Lambda \eta\pi^+)$ of Ref.~\cite{Duan:2024led}. This value could be tested by BESIII, Belle~II, and the proposed Super Tau-Charm Factory~\cite{Cheng:2022tog,Guo:2022kdi}  experiments in the future, which is crucial to understanding the role of the $a_0(980)$ in the process $\Lambda_c^+ \to \pi^+\eta n/\Lambda_c^+ \to \Lambda\eta\pi^+$.

%estimate the ratio $R$ = $\mathcal{B}(\Lambda_c^+ \to a_0(980)^+ n)/\mathcal{B}(\Lambda_c^+ \to \pi^+\eta n)\approx 0.313$.

%Finally, we could estimate the ratio $R$ = $\mathcal{B}(\Lambda_c^+ \to a_0(980)^{+} n)/\mathcal{B}(\Lambda_c^+ \to \pi^+\eta n)\approx 0.313$ based on the theoretical formula of Eq.~(\ref{eq:dgammadm12dm23}). 

%\begin{figure}[tbhp]
%	\centering
%	\includegraphics[scale=0.80]{Dali2-cut.pdf}
	%\vspace{-0.7cm}
%	\caption{The dalitz plot of “$M_{\pi^+\eta}$” vs “$M_{\eta n}$” for the process $\Lambda_c^+ \to \pi^+\eta n$ with a cut of $M_{\eta n}<1700$ MeV.}
%	\label{fig:dali-cut}
%\end{figure}

%$S$-wave pseudosalar meson-octet baryon interactions and $S$-wave pseudosalar-pseudosalar interactions within the chiral unitary approach

%%%%%%%%%%%%%%%%%%%%%%
\section{Summary} \label{sec:Conclusions}

The non-leptonic weak decays of charmed baryons are an important way to explore the properties of light hadrons. In this work, we have investigated the process $\Lambda_c^+ \to \pi^+\eta n$ taking into account the contributions from the $N(1535)$ and $a_0(980)$. We have assumed that $N(1535)$ is molecular state, which is dynamicallly generated by the $S$-wave pseudosalar meson-octet baryon interaction. Meanwhile, we also have considered the contribution of the scalar $a_0(980)$, which is dynamicallly generated by the $S$-wave pseudosalar-pseudosalar interaction. We have shown the results of the $\eta n$ and $\pi^+\eta$ invariant mass distributions, and found a near-threshold enhancement structure associated with $N(1535)$ and a cusp structure of the $a_{0}(980)$ in the $\eta n$ and $\pi^+\eta$ invariant mass distribution, respectively. Subsequently, we present the Dalitz plots of ``$M_{\eta n}^2$" vs. ``$M_{\pi^+ n}^2$", ``$M_{\pi^+\eta}^2$" vs. ``$M_{\eta n}^2$" and ``$M_{\pi^+\eta}^2$" vs. ``$M_{\pi^+ n}^2$" for the process $\Lambda_c^+ \to \pi^+\eta n$, respectively, where we can more clearly find the signals of $N(1535)$ and $a_0(980)$.

Then, in order to better display the contribution from the $a_0(980)$, we take a cut of $M_{\eta n}\geq1700$~MeV to eliminate the contribution from the $N(1535)$ and show the result of $\pi^+\eta$ invariant mass distribution. Meantime, we further estimate the ratio $R$ = $\mathcal{B}(\Lambda_c^+ \to a_0(980)^+ n)/\mathcal{B}(\Lambda_c^+ \to \pi^+\eta n)\approx 0.313$. These results can help us to better understand the strength and role of the scalar meson $a_0(980)$ in this process. 
Although this process has not been measured experimentally, its branching fraction without the resonant contribution is predicted to be $\mathcal{B}=(4.52\pm1.21)\times 10^{-3}$ in SU(3) flavor symmetry~\cite{Geng:2024sgq}.
Therefore, we strongly encourage our experimental colleagues, such as BESIII, Belle~II, and the proposed Super Tau-Charm Factory, to measure the $\Lambda_c^+ \to \pi^+\eta n$ process, which would be crucial to sheding light on the nature of the $N(1535)$ and $a_0(980)$.

%In the assumption of $N(1535)$ is molecular state, we have studied the process $\Lambda_c^+ \to \pi^+\eta n$ by considering the $S$-wave pseudosalar meson-octet baryon interaction, and the scalar $a_0(980)$ is dynamicallly generated by the $S$-wave pseudosalar-pseudosalar interactions. Then, we present the invariant mass distributions of the process $\Lambda_c^+ \to \pi^+\eta n$. There is a clear peak structure near 1535 MeV in the $\eta n$ invariant mass distribution, which corresponds to the $N(1535)$ resonance. In the $\pi^+\eta$ invariant mass distribution, we found a cusp structure near 980 MeV, which is associated with the $a_0(980)$. Furthermore, we also predict the Dalitz plot for the process $\Lambda_c^+ \to \pi^+\eta n$, and present the significant signals related to $a_0(980)$ and $N(1535)$.

%Since no measurements of the processes $\Lambda_c^+ \to \pi^+\eta n$ have been tested so far, we roughly calculate the ratio $R$ = $\mathcal{B}(\Lambda_c^+ \to a_0(980)^{+} n)/\mathcal{B}(\Lambda_c^+ \to \pi^+\eta n)\approx 0.313$. In the future, the BESIII and Belle II experiments will be able to verify our theoretical predictions, we expact experimentalists to measure this process to help us better comprehend the nature of $a_0(980)$ and $N(1535)$.

\section*{Acknowledgments}
This work is supported by the National Key R$\&$D Program of China (No. 2024YFE0105200), the Natural Science Foundation of Henan under Grant No. 232300421140 and No. 222300420554,  the National Natural Science Foundation of China under Grant No. 12475086, No. 12192263.

%\bibliographystyle{unsrt}
%\bibliography{ref}

\end{document}